 \documentclass[final,3p,times,sort&compress,12pt]{elsarticle}

\usepackage{amssymb}
\usepackage{hyperref}
\usepackage{breakurl,epstopdf,xcolor}

\journal{Carbon}

	\usepackage{fancyheadings}
\renewcommand{\thispagestyle}[1]{} 

\begin{document}

\begin{frontmatter}



\title{Critical temperature of two-dimensional hydrogenated multilayer graphene-based diluted ferromagnet}


\author{Karol Sza{\l}owski}

\address{Department of Solid State Physics, Faculty of Physics and Applied Informatics, University of {\L}\'{o}d\'{z},\\ ul. Pomorska 149/153, 90-236 {\L}\'{o}d\'{z}, Poland}

\date{\today}

 \cortext[cor1]{Corresponding author. E −mail address:
kszalowski@uni.lodz.pl}

\begin{abstract}
In the paper a theoretical study of critical (Curie) temperature of diluted ferromagnet based on multilayer graphene (or graphite) with hydrogen adatoms deposited over carbon atoms belonging to single sublattice is presented. The calculations are performed within Pair Approximation (PA) for diluted ferromagnetic system with long-range interactions. The method is able to take into account the spin-space anisotropy of coupling. The results obtained within Mean Field Approximation (MFA) are also presented for comparison. The assumed interaction between hydrogen adatom spins is inversely proportional to their mutual distance, with the additional exponential attenuation reflecting the presence of disorder in the system. The results obtained for a wide range of impurity concentrations and interaction decay length are discussed. The strongly non-linear behaviour of critical temperature as a function of dilution is predicted, at variance with MFA predictions. Moreover, MFA tends to overestimate heavily the critical temperature values compared to PA. An universal dependence of critical temperature on impurity concentration and interaction decay length is found for strong dilution regime.

\end{abstract}


\end{frontmatter}

\section{Introduction}

The search for novel magnetic materials is at the cutting edge of spintronic research. One of the highly promising, yet still challenging platforms for spintronic devices is graphene and various related systems \cite{Han2016,Gmitra2015,Han2014,Roche2014,Seneor2013,Pesin2012,Rycerz2007}. In particular, hydrogenated graphene attracts both the theoretical and experimental attention \cite{Soriano2015,Gmitra2014,Putz2014,Deng2014,Gmitra2013,Guo2013,Soriano2010,Zhou2009}. The natural goal to achieve is first the introduction of localized magnetic moments to graphene and secondly achieving the ferromagnetic ordering of the moments. Various paths towards ferromagnetic graphene systems have been proposed,  one of them being the creation of defects, including especially formation of vacancies and deposition of adatoms \cite{Jana2016, Yazyev2010}. One of the most fruitful approaches involves the light adatoms on graphene, to mention especially hydrogen \cite{Boukhvalov2008,Sofo2012}. The interactions between the chemisorbed adatoms and the graphene cause a localized spin equal to 1/2 to emerge. Such a phenomenon has been already demonstrated experimentally, both with the help of scanning tunneling microscopy methods \cite{Herrero2016} and in earlier direct magnetometric experiment \cite{Nair2012,Nair2013} as well as within some indirect approaches, utilizing the spin transport measurements \cite{McCreary2009,Hong2012}.

The existence of long-range coupling between magnetic impurities in graphene-based systems is predicted. For example, indirect, charge-carrier mediated Ruderman-Kittel-Kasuya-Yosida (RKKY) interaction has a long-range nature and has been extensively studied in monolayer graphene \cite{Dugaev2006,Annika2010b,Kogan2011}, but also in the multilayer case \cite{Jiang2012}. Its unique feature in graphene is the mutual relation between the coupling sign and the bipartite nature of graphene lattice: namely, the impurities located over the sites of the same sublattice should couple ferromagnetically \cite{Saremi2007}. The Density Functional Theory (DFT) was also employed to investigate the magnetic interactions of various adatoms on graphene surface, predicting usually a power law-like decay of coupling \cite{Pisani2008,Santos2010,Santos2012,Moaied2014}. In particular, Ref.~\cite{Moaied2014} predicts the ferromagnetic coupling falling off as $1/r$ between hydrogen adatoms on the surface of multilayer graphene (or graphite) if deposited in the same sublattice. Such interaction for graphene was also found in Ref.~\cite{Shytov2009}. The example of Ref.~\cite{Moaied2014} will be of special interest to us.

A key prediction related to the properties of a diluted ferromagnetic material is the critical (Curie) temperature and its dependence on the concentration of magnetic impurities. Making such a prediction requires the knowledge of the interactions between the spins as well as construction of the thermodynamic description of the magnetic system. Apart from the interest in the critical temperature of graphene-based magnet due to its applicational potential in spintronics, such systems are also interesting from the purely thermodynamic point of view. 

The aim of the paper is to present the theoretical study of the critical temperature of diluted, hydrogenated graphene-based ferromagnet with a method reaching beyond MFA. The system of interest consists of multilayer graphene (or graphite) with localized magnetic moments due to the presence of the hydrogen adatoms. The interaction between the moments is a modified coupling predicted in Ref.~\cite{Moaied2014}. Let us emphasize that predictions of Ref.~\cite{Moaied2014} concern graphite or graphene with at least five monolayers; therefore, we consider our system to be multilayer graphene of sufficient thickness. In the further sections we discuss in details the theoretical model of our system as well as we present extensively the numerical results.

\section{Theoretical model}

\begin{figure}[h!]
  \begin{center}
   \includegraphics[scale=0.82]{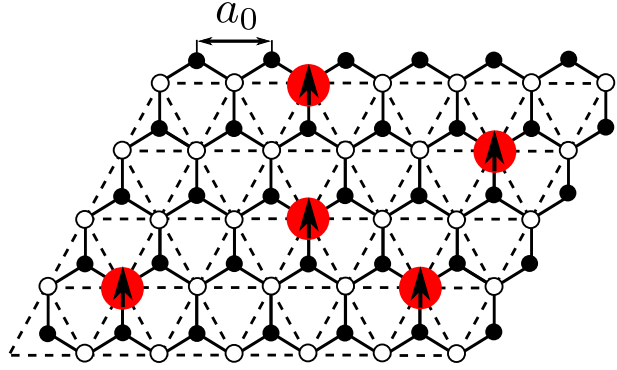}
  \end{center}
   \caption{\label{fig:scheme} Top view of the system of H adatoms randomly distributed over the sites of a single sublattice on graphite (multilayer graphene) surface. Only the topmost layer of C atoms is shown.}
\end{figure}

Our system of interest is a system of spin-1/2 magnetic impurities distributed over two-dimensional hexagonal lattice sites in on-site positions. Because we are interested in conditions when long-range ferromagnetic coupling between impurities can exist, we consider the particular case when the magnetic impurities are located only over the atoms belonging to one of two sublattices \cite{Moaied2014}. Therefore, the resulting underlying lattice for our diluted magnet is a triangular one (as illustrated in Fig.~\ref{fig:scheme} with dashed lines). The parameter $a_0=2.454$ \AA\ \cite{Moaied2014} is the distance between nearest neighbouring atoms in the same sublattice. We denote the concentration of impurities (i.e. H atoms) per lattice site of this triangular lattice by $p$. Therefore, the coverage of surface with the adatoms is equal to $p/2$. 

We assume that there exists a long-range ferromagnetic interaction between impurity spins, which takes the form of 
\begin{equation}
\label{eq:eq2}
J\left(r\right)=J_0\frac{a_0}{r}\exp\left(-r/\lambda\right),
\end{equation}
where for further numerical calculations we take the value of $J_0/k_{\rm B}=2.67\cdot 10^3$ K according to Ref.~\cite{Moaied2014}. The extensive discussion of the choice of interaction in such a form can be found in the Section \ref{Discussion}. Let us mention here that the parameter $\lambda$ is an exponential damping characteristic length which we treat as a phenomenological parameter reflecting the disorder. 

In a system of spins 1/2 with long-range ferromagnetic couplings, a presence of second-order phase transition can be expected. For the purpose of calculations of its critical temperature (Curie temperature) we apply the Pair Approximation method, which we have generalized in our earlier work \cite{Szalowski2014} to diluted ferromagnetic systems with long-range interactions. The method is superior to Mean-Field Approximation since it takes into account quantum fluctuations and is therefore capable of treating anisotropies in spin space (for example it distinguishes between Ising and Heisenberg coupling cases). 

We describe the diluted system of interacting magnetic impurities by means of the following Hamiltonian: 
\begin{equation}
\mathcal{H}=-\sum_{i,\,j}^{}{J\left(r_{ij}\right)\left[\Delta\left(S^{x}_{i}S^{x}_{j}+S^{y}_{i}S^{y}_{j}\right)+S^{z}_{i}S^{z}_{j}\right]\xi_{i}\xi_{j}},
\label{eq:eq1}
\end{equation}
where $J\left(r_{ij}\right)>0$ is the exchange integral of long-range ferromagnetic interactions between impurities of mutual distance equal to $r_{ij}$, as given by Eq.~\ref{eq:eq2}. The dilution is introduced by the operators $\xi_{i}$ which take the values of 1 when a magnetic adatom is present over the site $i$ and 0 otherwise. The parameter $\Delta$ describes the coupling anisotropy in spin space and can take arbitrary value ranging from 0 (Ising coupling) to 1 (isotropic Heisenberg coupling). As we have shown in the work Ref.~\cite{Szalowski2014}, the equation for critical temperature of continuous phase transition (Curie temperature) has the following form: 
\begin{equation}
p\sum_{k=1}^{\infty}{z_{k}\left[1-\exp\left(-\frac{1}{2}\beta_{\mathrm C}J_{k}\right)\cosh\left(\frac{1}{2}\beta_{\mathrm C}J_{k}\Delta\right)\right]}=2,
\label{eq:TC}
\end{equation}
where $\beta_{\mathrm C}=1/\left(k_{\rm B}T_{\mathrm C}\right)$, $T_{\mathrm C}$ is the critical temperature and $k_{\rm B}$ is Boltzmann constant. The summation extends over all the coordination zones of the studied lattice, characterized by radii $r_{k}$ and number of atoms (at the distance $r_{k}$ from the selected origin) equal to $z_{k}$, and $J_{k}\equiv J\left(r_{k}\right)$. The equation can be also written as:
\begin{eqnarray}
&&\sum_{k=1}^{\infty}{z_{k}\left\{1-\exp\left[-\frac{\beta_{\mathrm C}}{2}J_{k}\left(1+\Delta\right)\right]\right\}}\nonumber\\&&+\sum_{k=1}^{\infty}{z_{k}\left\{1-\exp\left[-\frac{\beta_{\mathrm C}}{2}J_{k}\left(1-\Delta\right)\right]\right\}}=\frac{4}{p}
\label{eq:TC2}
\end{eqnarray}
This equation can be solved only numerically for the case of long-range interactions.

Let us also present, for the purpose of comparison, the prediction of critical temperature within MFA, given by the formula:
\begin{equation}
k_{\rm B}T^{MFA}_{\mathrm C}=\frac{1}{4}p\sum_{k=1}^{\infty}{z_{k}J_{k}},
\label{eq:TCMFA}
\end{equation}
which is strictly linear in $p$.

When $T_{\mathrm C}\to\infty$, in equation \ref{eq:TC2} the approximation $\exp\left[-\frac{\beta_{\mathrm C}}{2}J_{k}\left(1\pm\Delta\right)\right]\simeq 1-\frac{\beta_{\mathrm C}}{2}J_{k}\left(1\pm\Delta\right)$ can be used. As a consequence, the equation is reduced exactly to the MFA expression given by Eq.~\ref{eq:TCMFA}. In particular, such a reduction takes place for $\lambda\to\infty$.

For low concentration $p$, the summation over lattice sites can be replaced with integration: \begin{equation}
\sum_{k=1}^{\infty}{z_{k}f\left(r_{k}\right)}\to \frac{2\pi}{\Omega_0}\,\int_{0}^{\infty}{r\,f\left(r\right)\,dr},
\label{eq:integral}
\end{equation} 
where the quantity $\Omega_0=\left(\sqrt{3}/2\right) \,a_0^2$ is the area per single sublattice site in graphene honeycomb lattice. Therefore, the equation \ref{eq:TC2} is transformed to:

\begin{eqnarray}
&&\!\!\!\!\!\!\!\!\!\!\!\!\!\!\!\!\frac{2\pi}{\Omega_0}\int_{0}^{\infty}{r\left\{1-\exp\left[-\frac{\beta_{\mathrm C}J_0a_0}{2}\frac{e^{-r/\lambda}}{r}\left(1+\Delta\right)\right]\right\}\,dr}\nonumber\\&&\!\!\!\!\!\!\!\!\!\!\!\!\!\!\!\!+\frac{2\pi}{\Omega_0}\int_{0}^{\infty}{r\left\{1-\exp\left[-\frac{\beta_{\mathrm C}J_0a_0}{2}\frac{e^{-r/\lambda}}{r}\left(1-\Delta\right)\right]\right\}\,dr}=\frac{4}{p}
\label{eq:int1}
\end{eqnarray}

The substitution $\rho=r/\lambda$ yields:
\begin{eqnarray}
&&\!\!\!\!\!\!\!\!\!\!\!\!\!\!\!\!\frac{2\pi}{\Omega_0}\int_{0}^{\infty}{\rho\left\{1-\exp\left[-\frac{\beta_{\mathrm C}a_0 J_0 }{2\lambda}\frac{e^{-\rho}}{\rho}\left(1+\Delta\right)\right]\right\}\,d\rho}\nonumber\\&&\!\!\!\!\!\!\!\!\!\!\!\!\!\!\!\!+\frac{2\pi}{\Omega_0}\int_{0}^{\infty}{\rho\left\{1-\exp\left[-\frac{\beta_{\mathrm C}a_0 J_0 }{2\lambda}\frac{e^{-\rho}}{\rho}\left(1-\Delta\right)\right]\right\}\,d\rho}=\frac{4}{\lambda^2p}
\label{eq:int2}
\end{eqnarray}

As a consequence, it can be observed that the left-hand side depends on the quantity $\lambda T_{\mathrm C}$, while the right-hand side on the quantity $\lambda^2 p$. Therefore, the equation \ref{eq:int2} constitutes an universal relation between $\lambda T_{\mathrm C}$ and $\lambda^2 p$, i.e. $\lambda T_{\mathrm C}=f\left(\lambda^2 p\right)$, what allows to draw an universal curve showing the dependence of these variables. Such an universal relation depends additionally only on the interaction anisotropy $\Delta$.

The replacement of summation with integration can be also performed for equation \ref{eq:TCMFA}, yielding $\sum_{k=1}^{\infty}{z_{k}J_{k}}\to \frac{2\pi}{\Omega_0}\int_{0}^{+\infty}{rJ\left(r\right)\,dr}=\frac{2\pi}{\Omega_0}a_0 J_0 \lambda$ and finally:
\begin{equation}
k_{\rm B}T^{MFA}_{\mathrm C}=\frac{\pi}{\sqrt{3}}\frac{J_0}{a_0}p\lambda.
\label{eq:TCMFAint}
\end{equation}

\section{Numerical results}

In order to predict the values of critical temperature for our system in question, we have performed extensive numerical calculations based on solving Eq.~\ref{eq:TC} for a wide range of concentrations $p$ and interaction decay parameters $\lambda$. Various interaction anisotropies $\Delta$ (ranging from Ising coupling to isotropic Heisenberg coupling) were included. Since the magnetic coupling in the system is a long-range one, we have used an appropriate cut-off $r_{max}$ for the radius of the coordination zones over which the summation in Eq.~\ref{eq:TC} takes place. We found that the results are convergent to the limit of infinite interaction range when $r_{max}\simeq 15\lambda$, which value we have used in the calculations. For the purpose of comparison, also the MFA approach was used, yielding the results based on Eq.~\ref{eq:TCMFA}. In the present section, the obtained numerical results will be discussed.

\begin{figure}[h!]
  \begin{center}
   \includegraphics[scale=0.32]{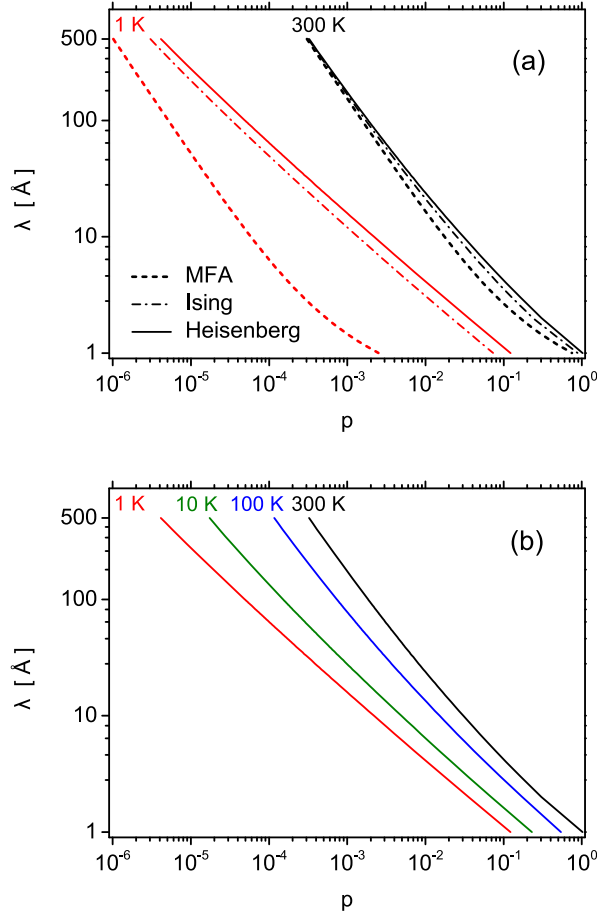}
  \end{center}
   \caption{\label{fig:isolines1} Lines of constant critical temperature plotted against concentration of H adatoms and interaction decay length. (a) the comparison of results yielded by MFA and by PA for Heisenberg and Ising interactions for two critical temperatures: 1K and 300 K; (b) the results of PA for Heisenberg interactions for four critical temperatures: 1K, 10 K, 100K and 300 K.}
\end{figure}

Let us commence the discussion from the prediction what values of concentration $p$ and interaction decay parameter $\lambda$ are necessary to achieve particular values of critical temperature. In order to compare the results yielded by MFA and PA (for Ising and Heisenberg couplings) for a wide range of $p$ and $\lambda$, we plot the lines of constant critical temperature (critical isotherms) as a function of both mentioned variables. Such a plot is shown in Fig.~\ref{fig:isolines1}(a) for two values of critical temperature: 1 K and room temperature equal to 300 K, in double logarithmic scale. It is evident that for the case of the lower temperature, the discrepancy between MFA and PA result is pronounced, reaching orders of magnitude, even for the case of $\lambda$ reaching 500 \AA, where the difference is still significant. In the scale of the plot, the influence of interaction anisotropy in spin space, as captured by PA, is less pronounced (note the double logarithmic scale). For the case of room temperature predictions, both methods yield more coherent results, in particular for weak interaction damping (large $\lambda$). However, at higher concentrations of H atoms and more localized charge carriers (which two factors may be mutually correlated in disordered system) there is still an important difference between MFA and PA.

For the case of Heisenberg coupling (which is expected in the absence of additional sources of anisotropy), we present the predictions of PA for critical isotherms in Fig.~\ref{fig:isolines1}(b) for wider range of critical temperatures: 1 K, 10 K, 100 K and 300 K. The lowest temperature critical isotherm is almost linear, while at higher temperatures the isotherms develop more convex shape. The room temperature is reached at the concentration $p$ approximately equal to $3\cdot10^{-4}$ for $\lambda=500$ \AA. 

\begin{figure}[h!]
  \begin{center}
   \includegraphics[scale=0.32]{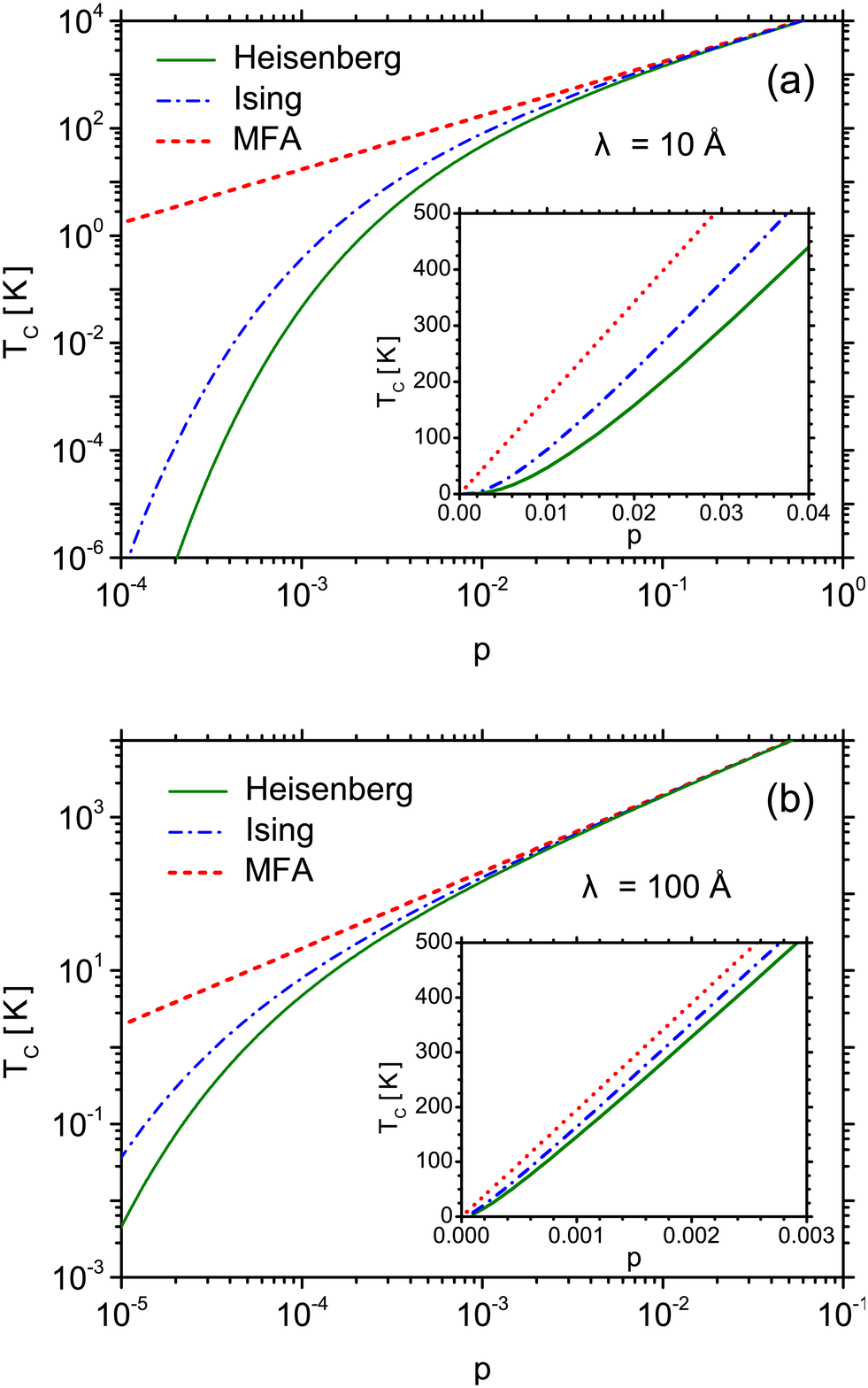}
  \end{center}
   \caption{\label{fig:tcvsp} Dependence of critical temperature on concentration of H atoms for two interaction decay lengths: 10 \AA\ (a) and 100 \AA\ (b). The results of MFA are compared with the results of PA for Heisenberg and Ising interactions. The insets present the same data in linear scale, for the range emphasizing the vicinity of the room temperature. }
\end{figure}

The most interesting results are connected with the form of the critical temperature dependence on concentration of magnetic moments. It is worth recalling that MFA predicts always a linear dependence, regardless of the spatial dependence of long-range coupling. The critical temperature is plotted as a function of concentration $p$ in Fig.~\ref{fig:tcvsp} in double logarithmic scale, due to significant dynamics of the data. Two cases are distinguished: Fig.~\ref{fig:tcvsp}(a) is prepared for $\lambda=10$ \AA, so relatively strongly damped coupling, while Fig.~\ref{fig:tcvsp}(b) assumes much longer $\lambda=100$ \AA. For both cases a huge overestimation of the critical temperature by MFA is notable unless the concentration is very high. The function $T_{\mathrm C}(p)$ predicted by PA is highly non-linear. Moreover, the method allows to take into account the interaction anisotropy in spin space, what is reflected in the difference between the Ising and isotropic Heisenberg coupling case. This difference tends to vanish (in the double logarithmic scale) at sufficiently high values of concentration $p$, which values decrease with decreasing interaction decay length. For sufficiently high concentration, both the results of PA and MFA tend to overlap (especially for higher $\lambda$), reflecting the limiting case described by Eq.~\ref{eq:TCMFA}. However, the regime of linear dependency of $T_{\mathrm C}$ on $p$ is reached, even for $\lambda =100$ \AA, at considerably high concentrations of hydrogen impurity. Therefore, the remarkable feature of the physically interesting range of concentrations is the manifestly non-linear rise of critical temperature with increasing $p$. The insets in Fig.~\ref{fig:tcvsp} present the critical temperature as a function of concentration in linear scale, covering the range of room temperatures. The differences between MFA results and the outcome of PA for limiting cases of interaction anisotropy are the most pronounced at low value of $\lambda=10$ \AA\ (inset in Fig.~\ref{fig:tcvsp}(a)). However, even at $\lambda=100$ \AA\ still the difference remains notable (as seen in inset in Fig.~\ref{fig:tcvsp}(b)). 

\begin{figure}[h!]
  \begin{center}
   \includegraphics[scale=0.32]{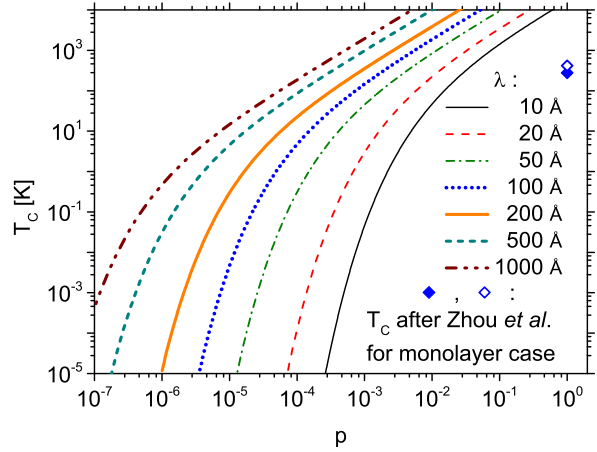}
  \end{center}
   \caption{\label{fig:temps} Dependence of critical temperature on H adatoms concentration predicted by PA for Heisenberg interactions for a wide range of interaction decay lengths: from 10 \AA\ to 1000 \AA. The filled and open diamonds denote the predictions of the critical temperature for half-hydrogenated monolayer graphene taken from Ref.~\cite{Zhou2009}: open symbol for 3D result, filled symbol for 2D one.}
\end{figure}

The dependence of the critical temperature on concentration of H for the case of Heisenberg couplings is shown in Fig.~\ref{fig:temps}. A wide range of interaction damping parameter $\lambda$ is covered, i.e. $\lambda=10,\dots,1000$ \AA. It is evident that the dependence is non-linear at lower concentrations and becomes linearized for higher ones. The linearity regime extends towards lower concentrations when $\lambda$ increases. The room critical temperature is promised to be reached at the concentrations of a few times $10^{-4}$ at the weakest interaction damping, $\lambda=1000$ \AA. However, it should be emphasized that achieving such a slow interaction decay can be a challenge, as stated further in the Section \ref{Discussion-2}.

In Fig.~\ref{fig:temps} we have marked, as a reference, the predictions of Ref.~\cite{Zhou2009} for critical temperature of half-hydrogenated monolayer graphene (i.e. with fully hydrogenated single sublattice), what corresponds to the limit of $p=1$ in our model (but was calculated for a monolayer). It is clearly visible that the predicted values are certainly well below our predictions. It should be, however, emphasized that our considerations concern the case of strongly diluted magnet. The critical temperature value of Ref.~\cite{Zhou2009} proves that the magnetic interactions for the fully hydrogenated sublattice in monolayer graphene appear weaker than for more diluted magnet based on multilayer graphene. This situation calls for a DFT calculation for the multilayer case with the highest possible hydrogenation for comparison.  

\begin{figure}[h!]
  \begin{center}
   \includegraphics[scale=0.32]{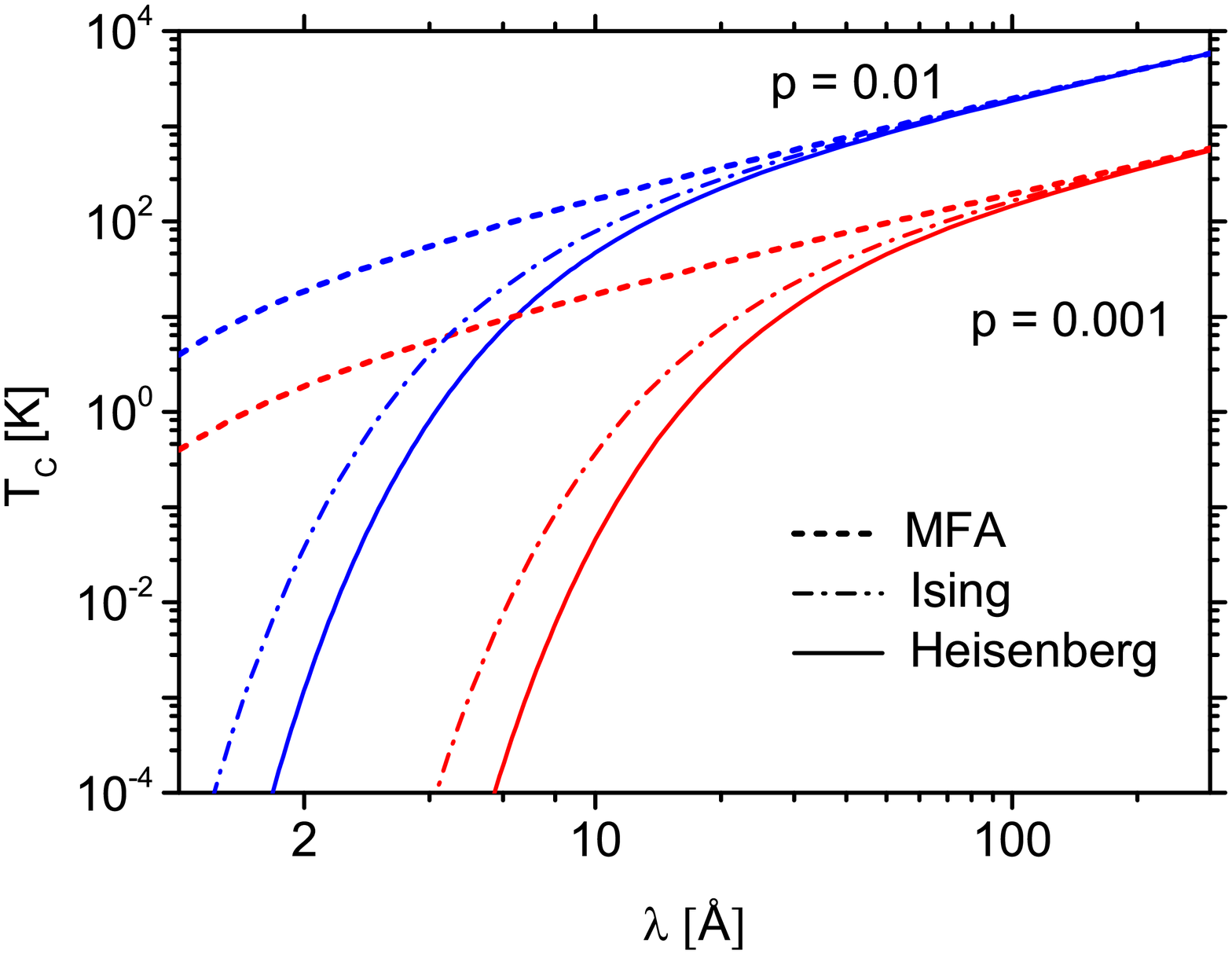}
  \end{center}
   \caption{\label{fig:tcvslambda} Dependence of critical temperature on interaction decay length for two concentrations of H atoms: 0.01 and 0.001, as predicted by MFA and by PA for Heisenberg and for Ising interactions.}
\end{figure}

The dependence of critical temperature on coupling decay parameter $\lambda$ is shown in double logarithmic scale in Fig.~\ref{fig:tcvslambda}, for two impurity concentrations: $p=10^{-2}$ and $p=10^{-3}$. Again, it can be concluded that the difference between MFA and PA results is significant unless the interaction damping is very weak (what situates the system in the range of validity of the MFA, see Eq.~\ref{eq:TCMFA}. This range is reached already for smaller $\lambda$ values if the concentration of magnetic impurities is larger, thus emphasizing the importance of non-MFA calculations for predicting the critical temperature for strongly diluted systems. The critical temperature varies non-linearly with $\lambda$ for such case and gradually acquires linear behaviour if the damping becomes weak enough.

\begin{figure}[h!]
  \begin{center}
   \includegraphics[scale=0.32]{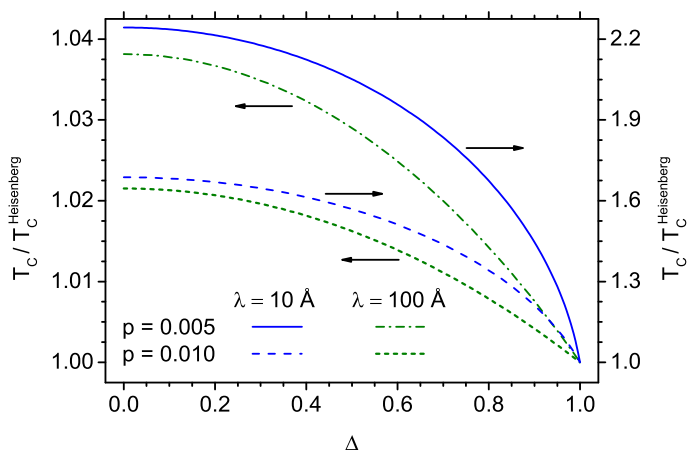}
  \end{center}
   \caption{\label{fig:anisotropy} Dependence of critical temperature normalized to critical temperature for Heisenberg interactions on interaction anisotropy parameter $\Delta$. Two values of interaction decay length: 10 \AA\ (left axis scale) and 100 \AA\ (right axis scale) and two values of H atoms concentration: 0.005 and 0.010 are considered. Please note the left and right axis scale difference.}
\end{figure}

As mentioned, PA is able to reflect the effect of spin space anisotropy on the critical temperature, to which anisotropy the MFA method is insensitive. Therefore, it can be useful to take a closer look at the relative importance of the anisotropy for the value of critical temperature for various ranges of parameters. The measure of the anisotropy is the ratio $\Delta$, taking the extreme values of 0 for Ising coupling and 1 for isotropic Heisenberg coupling. The relative critical temperature (related to its value at $\Delta =1$, i.e. for isotropic Heisenberg coupling) is plotted as a function of $\Delta$ in Fig.~\ref{fig:anisotropy}. The left vertical scale is valid for $\lambda=100$ \AA, while the right one - for $\lambda=10$ \AA. In general, coupling anisotropy causes the critical temperature to increase (for example, for ferromagnets with nearest-neighbour interactions only, the critical temperature for Ising case is higher than for Heisenberg case). However, it can be easily observed that the relative change in critical temperature is significantly larger for stronger damping of long-range coupling. For $\lambda=100$ \AA\ it reaches at best a few per cent, while for $\lambda=10$ \AA\  it can be of the order of a hundred per cent. Therefore, it can be emphasized that for weak interaction damping, which paves the way towards high critical temperatures, the influence of coupling anisotropy in spin space would not be a significant factor leading to further increase of $T_{\mathrm C}$. This statement is further supported by the fact that the relative change in critical temperature decreases with rising impurity concentration, which is another route to maximizing the critical temperature. As a consequence, inducing the coupling anisotropy in spin space would be an effective way of increasing the Curie temperature only in the range of parameters where the temperature would be itself low. In the range of validity of Eq.~\ref{eq:TCMFA}, $T_{\mathrm C}$ completely loses its sensitivity to coupling anisotropy and the system exhibits mean-field behaviour.

\begin{figure}[h!]
  \begin{center}
   \includegraphics[scale=0.32]{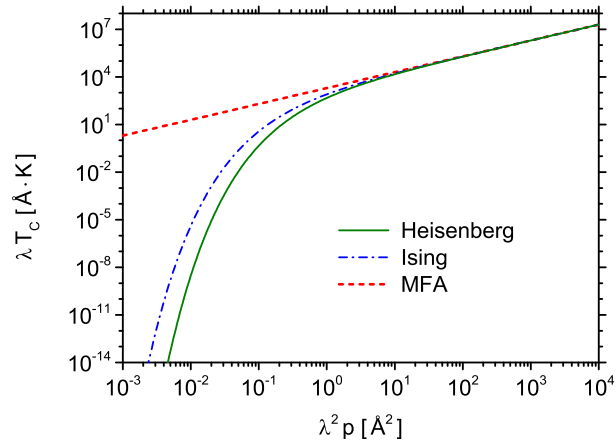}
  \end{center}
   \caption{\label{fig:tcuniversal} Universal dependence of $\lambda T_{\mathrm C}$ on $\lambda^2 p$ predicted by MFA and by PA for Heisenberg and for Ising interactions, when the integration over lattice is used (see Eq.~\ref{eq:int2} for PA and Eq.~\ref{eq:TCMFAint} for MFA). }
\end{figure}

As it was mentioned in the previous section, it can be shown that the critical temperatures fall onto an universal curve provided that the the parameters $p$ and $\lambda$ are within the validity range of integral approximation Eq.~\ref{eq:integral}. In such a range, the universal function $\lambda T_{\mathrm C}=f\left(\lambda^2 p\right)$ exists and its form depends only on the spin-space anisotropy parameter $\Delta$.  Such an universal dependence is demonstrated in Fig.~\ref{fig:tcuniversal}, for the limiting cases of Ising and Heisenberg interactions, based on Eq.~\ref{eq:int2}. Also the MFA dependence (Eq.~\ref{eq:TCMFAint}) is shown for comparison. Let us note that, in general, the integral approximation is valid for low $p$ (see also Refs.~\cite{Szalowski2014,Fabritius2010}). 

It is certainly worth emphasizing that the existence of an universal relation of $\lambda T_{\mathrm C}$ and $\lambda^2 p$ can be utilized to validate the model based on the exponential damping of the magnetic interaction on the grounds of experimental data. On the other hand, it could also provide a route towards determination of the attenuation length $\lambda$ from measurements of the critical temperature.

\section{Discussion}
\label{Discussion}

\subsection{The magnetic impurities}

Let us begin the discussion from the remarks concerning the validity of our model for various concentrations of hydrogen atoms $p$. The scope of our model involves the diluted graphene ferromagnet. Therefore, the concentration of adatoms should be low enough to guarantee that their average separation exceeds the in-plane radius of magnetic moment induced by each hydrogen adatom. On the basis of the recent experimental study \cite{Herrero2016}, we roughly estimate that the corresponding radius $r_0$ is of the order of 10 \AA. Therefore, for the validity of the model, the inequality $\Omega_0/p\gtrsim \pi r_0^2$ should be satisfied, leading roughly to $p\lesssim 0.01$. As a consequence, the results for $p\gtrsim 0.01$, presented for completeness, require careful approach, with the awareness that in such range the interaction of magnetic moments can be more complicated than we assume. 

Let us strongly emphasize that we are interested only in the case when the hydrogen is deposited in a form of single, isolated adatoms, which type of defect leads to the formation of spin equal to 1/2. On the contrary, pairs or larger clusters tend to be either non-magnetic or at least exhibit significantly reduced magnetic moment in relation to the number of H atoms, as shown in DFT calculations \cite{Ranjbar2010,Cao2013,Alam2013}. For example, non-magnetic dimers are found to be the most energetically stable dimers \cite{Ranjbar2010}. Such effect was also suspected to cause the important discrepancy between total observed magnetic moment and concentration of adatoms in the experiment \cite{Nair2012}. The probability of creation of the clusters increases with increasing $p$, so that our main range of interest is the regime of strong dilution, far from strong hydrogenation, that leads inevitably to formation of clusters and finally to graphone. 

It should be also mentioned that the ferromagnetic ordering of adatoms deposited over the sites of the same sublattice due to ferromagnetic couplings should decrease the total energy at the same time preferring such distribution of impurities in comparison with their deposition over both sublattices \cite{Cheianov2010}. Also other factors may contribute to that effect \cite{Talbot2016}, stabilizing the system in the state with the impurities occupying just one sublattice. On the other hand, this means that the distribution of adatoms is not of purely random character and their positions can exhibit some correlations. The presence of such correlations would influence the thermodynamic description of the system (as demonstrated, for example, for the case of diluted magnets with RKKY interaction \cite{Szalowski2008} or with interaction between nearest neighbours \cite{Balcerzak2009}).

Such factors as correlated distribution of hydrogen impurities or growing tendency for charge carriers localization with the increase of impurity concentration are certainly worth incorporating into the detailed theory of magnetism in hydrogenated graphene systems. Therefore, we believe that our calculations can serve as a useful starting point for further developments in this field, aimed at achieving robust carbon-based ferromagnetic materials.

An important question regarding the stability of the considered system can also be the issue of mobility of the adatoms on the surface. However, a suppression of hopping of impurities can be expected according to Ref.~\cite{Moaied2015}, what suggests the decent stability of the system. 

\subsection{The magnetic long-range interaction}
\label{Discussion-2}

Let us now comment on the selection of the interaction between magnetic impurities in the form of $J\left(r\right)\propto (1/r)\exp\left(-r/\lambda\right)$. Among the DFT-based studies of interspin coupling for particular kinds of magnetic moments in graphene, we notice for example the prediction of power law-like decay of coupling between magnetic defects for periodically arranged hydrogenated vacancies in graphene \cite{Pisani2008}. A similar behaviour was found in Ref.~\cite{Santos2010} for Co adatoms. Calculations for a wider group of adsorbates \cite{Santos2012} also predict a power decay of ferromagnetic coupling for impurities in the same sublattice. It has been predicted in Ref.~\cite{Moaied2014} that the coupling between two H adatoms in the same sublattice is inversely proportional to their distance, that is $J\left(r\right)\propto 1/r$. 

We comment here that it might be difficult to predict the value of $\lambda$ and to observe its influence in calculations based on just a pair of impurities. This might be even more complex for $\lambda$ exceeding significantly the size of the used supercell. We emphasize the fact that the real system is strongly disordered because a large number of hydrogen impurities is present, while in DFT calculations we deal with a single pair of impurities. This strong disorder due to randomly distributed hydrogen impurities influences the electronic wavefunctions by causing the tendency to localization and, therefore, may limit the range of interaction. Such a phenomenon is known for example from the studies of indirect RKKY coupling in disordered systems \cite{deGennes1962}, including the case of indirect coupling in graphene \cite{Lee2012,Lee2012b}. The conclusion can be drawn that the disorder causes damping of amplitude of indirect interaction and also shifts the phase \cite{deGennes1962}. The first effect appears to be of larger importance and can be often approximated by means of introducing an exponential damping of the long-range interaction, i.e. the factor $\exp\left(-r/\lambda\right)$ (being an approximation, as pointed out in Ref.~\cite{deGennes1962}). Such a modification of coupling was used, for example, in calculations regarding the disordered diluted magnetic semiconductors (e.g. \cite{Story1992,Szalowski2008}). In the case of RKKY interaction damping in systems such as diluted magnetic semiconductors, $\lambda$ could be identified with mean free path of the charge carriers \cite{Kilanski2012,Kilanski2010}. However, for the system considered here, the interpretation of this parameter could be more sophisticated and less direct. The reason for introducing it is the necessity of including the effect of disorder on indirect interactions. We treat the introduction of an exponential damping factor as a phenomenological way of taking into account the inevitable presence of disorder in our diluted magnetic system. In addition to hydrogen adatoms, also other adatoms might be present on the surface. Also some local inhomogeneities of charge carrier concentration, in the form of the charge puddles might be suspected to influence the range of magnetic coupling. As a consequence, achieving very long damping lengths $\lambda$ would be rather challenging.    

In principle, the effective decay length of indirect coupling $\lambda$ and the concentration of magnetic impurities $p$ are not independent variables. On the contrary, the presence of impurities promotes the localization of charge carriers, decreasing the localization length (see for example the theoretical study in Ref.~ \cite{Gargiulo2014}, where the localization length for electronic wavefunctions was decreasing linearly with a logarithm of impurity concentration, as shown in Fig.~5(b) in that reference). Such effects might not be directly visible in any DFT calculations of indirect magnetic exchange based on a pair of impurities. In a real, multi-impurity system the distance dependence of the coupling might, therefore, differ from the two-impurity prediction. Therefore, it can be particularly challenging to achieve at the same time rather high concentration of impurities and weak damping of interaction.

\subsection{The PA method}

The description of magnetic systems with long-range interactions is usually performed on the grounds of MFA, which always predicts linear dependence of critical temperature on concentration of magnetic component. In such context, it is noteworthy that some graphene-based systems have been a subject of calculations by Monte Carlo method \cite{Fabritius2010,Qi2013,Masrour2015}. In particular, a diluted antiferromagnet has been investigated \cite{Fabritius2010}, revealing the finite-temperature ordering despite noticeably fast decay of interaction with the distance and low dimensionality. The critical temperature was also found to depend non-linearly on the concentration of magnetic moments. On the other hand, the studies of strongly diluted magnets with simulational methods require huge clusters to guarantee sufficient number of spins in the simulated system when interspin distances are large. This feature makes the simulations increasingly computationally demanding for decreasing concentration of magnetic moments. Therefore, studies employing more advanced methods than MFA, but of analytic character are well motivated. 

Let us mention here that PA method also provides qualitatively better results than MFA for a range of other physically interesting systems. In connection with the reliability of the PA, we found that it predicts the critical temperature proportional to $p^{-n/d}$ in a strongly diluted $d$-dimensional system with interactions falling off like $J\left(r\right)\propto 1/r^n$ \cite{Szalowski2014}. This result is in accord with qualitatively exact scaling predictions, contrary to MFA which yields $T_{\mathrm C}\propto p$ regardless of the distance dependence of interactions. Such a non-linear concentration dependence of the critical temperature has been experimentally observed, for example, in diluted magnetic semiconductor (Ga,Mn)N with long-range ferromagnetic superexchange interactions \cite{Dietl2014,Simserides2014,Stefanowicz2013,Sawicki2012}. For such systems, also Monte Carlo calculations were performed, predicting the already mentioned kind of non-linear dependence \cite{Dietl2014,Simserides2014,Stefanowicz2013,Sawicki2012}. Therefore, the PA method is clearly superior to MFA, remaining a non-simulational approach.

Let us state that PA is the method which provides the best results for all-ferromagnetic or al least dominantly ferromagnetic interactions in the system. However, it was also applied for example to the bilayer system with ferromagnetic intra- and antiferromagnetic interlayer interactions \cite{Balcerzak2014}. Therefore, it might be useful in the case of graphene system with both sublattices hydrogenated (when the coupling is ferromagnetic between the adatoms located at the same sublattice and antiferromagnetic otherwise). However, study of such a system would call for quite extensive reformulation of the model, showing the direction for future proceedings.

\section{Final remarks}

In the paper, we have presented the calculations of the critical temperature for ferromagnetic system consisting of multilayer graphene with partial hydrogenation of a single sublattice. We based the predictions on non-simulational Pair Approximation method, which is able to take into account spin-space anisotropy of interactions, thus it us superior to commonly applied Mean Field Approximation. We have used the interaction between adatoms predicted by Ref.~\cite{Moaied2014} modifying it by introducing the exponential damping due to disorder. The main parameters of our model are: the concentration of hydrogen adatoms per carbon lattice site in single sublattice and the interaction decay length. The results of extensive calculations of critical temperature as a function of these two parameters were presented and discussed.

Our main result is a non-linear dependence of the critical temperature on the impurity concentration. We have found that, in general, MFA tends to overestimate heavily the critical temperature, unless the damping of interaction is rather weak. Analysis of the influence of spin-space anisotropy on the critical temperature yields the conclusion that transition from isotropic Heisenberg to anisotropic Ising interaction can very effectively increase the critical temperature only for quite short damping lengths (where the temperature itself is rather low), but has very limited effect in the range where the critical temperature is already high. 

In the range of low concentrations of hydrogen, we found an universal dependence of $\lambda T_{\mathrm C}$ on $\lambda^2 p$. The form of the mentioned dependence is only influenced by the spin-space interaction anisotropy and might be useful for verification of the model vs. experimental data. 

We have also constructed critical isotherms in the space of two parameters: impurity concentration and interaction decay length, for various values of critical temperature. We believe that there results can be particularly useful for making predictions concerning the requirements to achieve the desired critical temperatures.

\section*{Acknowledgments}

\noindent The computational support on Hugo cluster at Laboratory of Theoretical Aspects of Quantum Magnetism and Statistical Physics, P. J. \v{S}af\'{a}rik University in Ko\v{s}ice is gratefully acknowledged.

\noindent This work has been supported by Polish Ministry of Science and Higher Education on a special purpose grant to fund the research and development activities and tasks associated with them, serving the development of young
scientists and doctoral students.

\section*{References}

\bibliographystyle{elsarticle-num}

\end{document}